\documentclass[epsfig,aps,prb,twocolumn,superscriptaddress,footinbib,amsmath,amssymb]{revtex4}

\usepackage{graphicx}
\usepackage{dcolumn}
\usepackage{bm}

\begin{document}

\title{Dynamical ultrafast all-optical switching of planar GaAs/AlAs photonic microcavities}
\author{Philip J. Harding}
\affiliation{Center for Nanophotonics, FOM Institute for Atomic and
Molecular Physics (AMOLF), Kruislaan 407, 1098 SJ Amsterdam,
The Netherlands}%
\author{Tijmen G. Euser}
\affiliation{Center for Nanophotonics, FOM Institute for Atomic and
Molecular Physics (AMOLF), Kruislaan 407, 1098 SJ Amsterdam,
The Netherlands}%
\author{Yoanna-Reine Nowicki-Bringuier}
\author{Jean-Michel G\'erard}
\affiliation{CEA/DRFMC/SP2M, Nanophysics and Semiconductor
Laboratory, 17 rue des Martyrs, 38054 Grenoble Cedex, France}
\author{Willem L. Vos}
\affiliation{Center for Nanophotonics, FOM Institute for Atomic and
Molecular Physics (AMOLF), Kruislaan 407, 1098 SJ Amsterdam,
The Netherlands}%
\affiliation{Complex Photonic Systems (COPS), MESA+ Institute for
Nanotechnology, University of Twente, The Netherlands}

\date{Prepared September 11th 2007}

\begin{abstract}
We study the ultrafast switching-on and -off of planar GaAs/AlAs
microcavities. Up to $0.8 \%$ refractive index changes are
achieved by optically exciting free carriers at $\lambda = 1720$
nm and pulse energy $E_{\rm{Pump}} = 1.8 \pm 0.18$ $\mu$J. The
cavity resonance is dynamically tracked by measuring reflectivity
versus time delay with tunable laser pulses, and is found to shift
by as much as 3.3 linewidths within a few ps. The switching-off
occurs with a decay time of $\sim50$ ps. We derive the dynamic
behavior of the carrier density, and of the complex refractive
index. We propose that our inferred 10 GHz switching rate may be
tenfold improved by optimized sample growth.
\end{abstract}

\maketitle

There is generally a great interest to store photons in a small
volume. This feat can be achieved in solid state structures with
tiny cavities, with dimensions of the order of the wavelength of
light. Light is so strongly confined in such cavities that large
electric field enhancements occurs. This field enhancement notably
leads to large modifications of the emission rate of an elementary
light source embedded inside a cavity \cite{Purcell:46,
Gerardetal:98}. It is highly desirable, both from fundamental and
applied viewpoints, to switch the optical properties of cavities
on ultrafast time scales. This ultrafast switching of cavities
will allow the catching or releasing of photons, changing the
frequency and bandwidth of confined photons, and even the
switching-on or -off of light sources, see Refs.\cite{Johnson:02,
Bret:03, Notomi:06, Fushman:07, Preble:07}. It is therefore
important to systematically study the dynamic behavior of switched
cavities. Surprisingly, such studies are scarce. Recently, Almeida
\textit{et al.} studied relaxation at two frequencies for a large
10 micron diameter Si ring resonator, revealing decay times of
0.45 ns \cite{Almeida:04}. Here, we use broadband tunable
femtosecond pump-probe reflectivity to study the dynamics of
planar thin $\lambda$-microcavities made from III-V
semiconductors, an important class of solid-state cavities that
are notably used in VCSELs \cite{Hense:97}.

Our sample consists of a GaAs $\lambda$-cavity with a thickness of
277 nm. The layer is sandwiched between two Bragg stacks
consisting of 12 and 16 pairs of $\lambda/4$ thick layers of
nominally pure GaAs or AlAs. The sample is grown with molecular
beam epitaxy at 550$^{0}$C to optimize the optical quality. A
slight variation of a few $\%$ in stopgap and cavity resonance
over the sample allowed us to verify the experimental observations
for different resonance wavelengths. For experiments outside the
present scope the sample was doped with $10^{10}$cm$^{-2}$
InGaAs/GaAs quantum dots, which hardly influence our experiment
\cite{noteQD}.

Our setup consists of two independently tunable optical parametric
amplifiers (OPA, Topas), that are the sources of the pump and probe
beams. The OPAs have pulse durations of $\tau_{\rm{P}} = 140 \pm 10$
fs corresponding to a spectral width of $1.4 \%$. The pump beam has
a much larger Gaussian focus of 113 $\mu$m FWHM than the probe beam
(28 $\mu$m), ensuring that only the central flat part of the pump
focus is probed. Free carriers are excited in the GaAs by two-photon
absorption at $\lambda = 1720$ nm, to obtain a spatially homogeneous
distribution of carriers \cite{Euser:05}. A versatile measurement
scheme was developed to subtract the pump background from the probe
signal, and to compensate for possible pulse-to-pulse variations in
the output of our laser \cite{Euseretal:06, Euser:07a}.
Continuous-wave (cw) reflectivity was measured with a broad band
white light setup with a resolution of $\sim 0.2$ nm.

\begin{figure}
\begin{center}
\includegraphics[angle=-90, width=8.0cm]{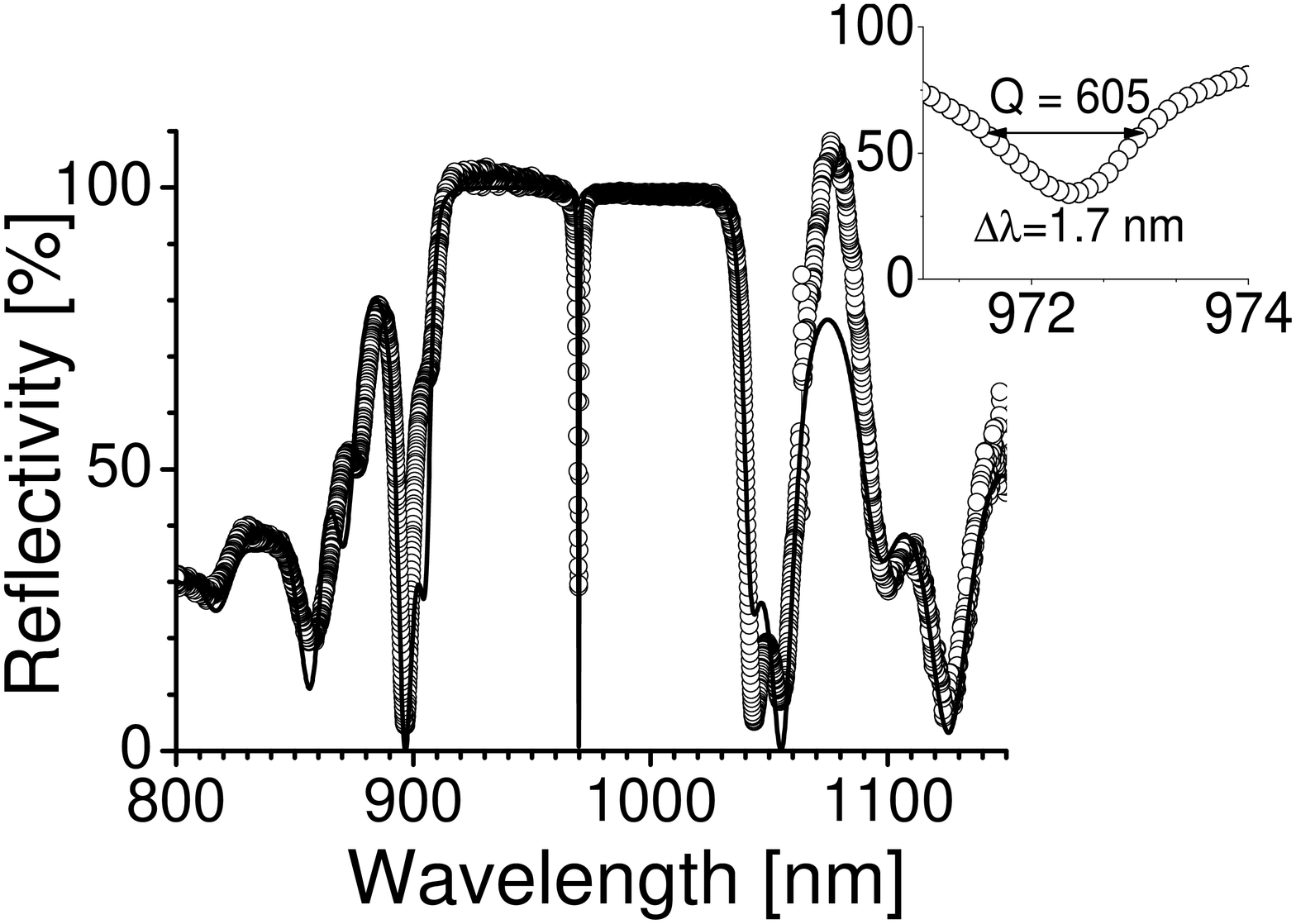}
\caption{Continuous wave reflectivity spectrum (open circles) at
normal incidence of sample with a resolution of 0.2 nm. The
resonance of the $\lambda$-cavity (Q = 605) can clearly be seen at
972 nm (see inset). The solid curve is a transfer matrix calculation
that includes the dispersion and absorptionof GaAs and AlAs.}
\label{fig:Refl}
\end{center}
\end{figure}

Fig. \ref{fig:Refl} shows a cw reflectivity spectrum of the planar
photonic microcavity at normal incidence. The high peak between 900
and 1040 nm is due to the stopgap of the Bragg stacks. The stopband
has a broad width $\Delta \lambda$ = 140 nm ($14.3 \%$ relative
bandwidth), which confirms the high photonic strength. The
Fabry-P\'{e}rot fringe at 1080 nm exceeding 100 $\%$ is due to some
chromatic abberation in the focus. Near 970 nm we observe a sharp
resonance caused by the $\lambda$-cavity in the structure. The
resonance has a linewidth $\Delta \lambda_{\rm{cavity}} = 1.7$ nm
(see inset), corresponding to a quality factor $Q = 605$. A transfer
matrix (TM) calculation including the dispersion and absorption of
GaAs \cite{Blakemore:82} and AlAs \cite{Fern:71} reproduces the
experimental resonance, stopband, and Fabry-P\'{e}rot fringes. The
only free parameters in the model were the thicknesses of the GaAs
($d_{\rm{GaAs}} = 68.78$ nm) and AlAs ($d_{\rm{AlAs}} = 81.90$ nm).

\begin{figure}
\begin{center}
\includegraphics[angle=-90, width=8.0cm]{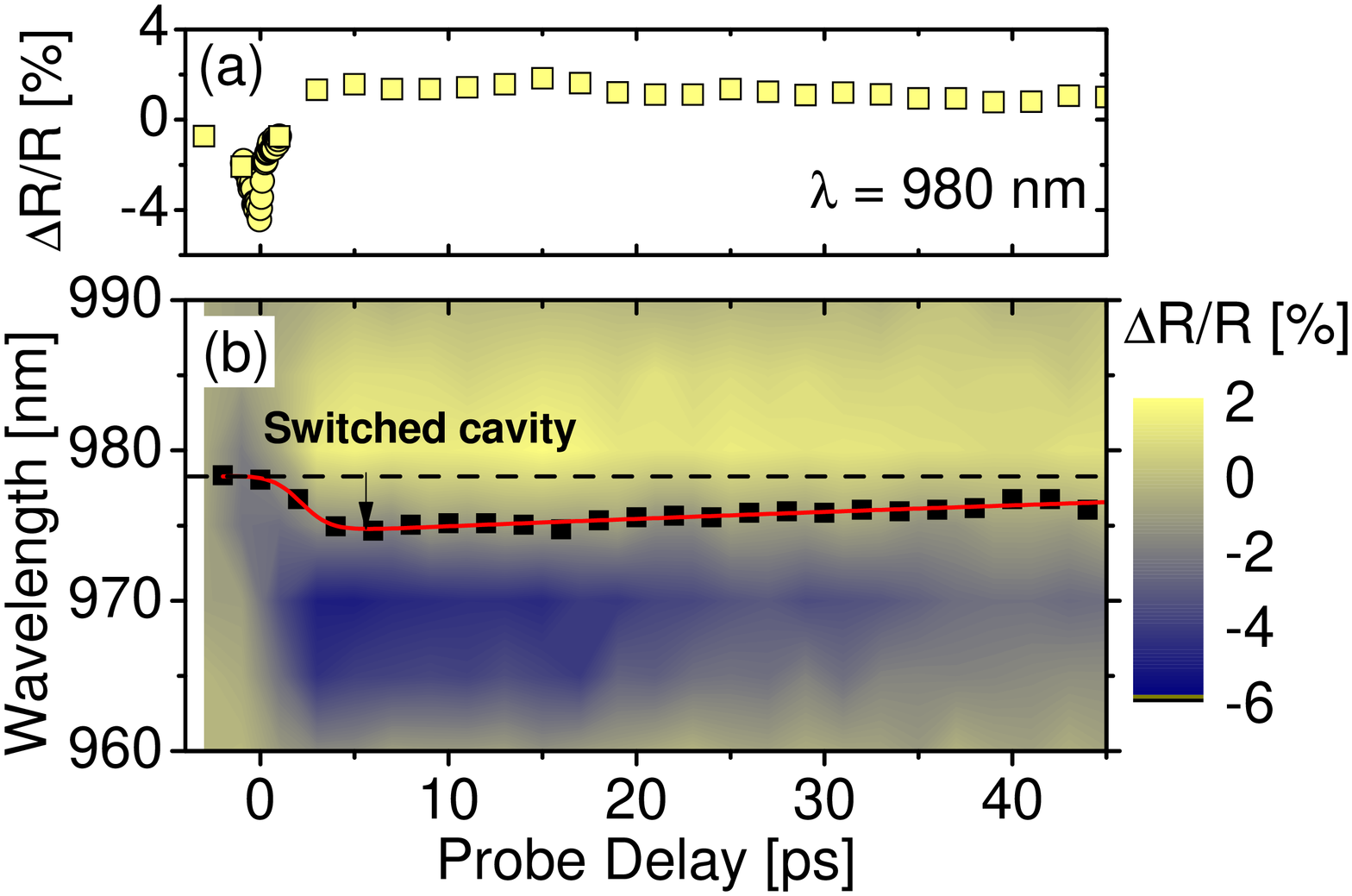}
\caption{(Color online) (a) Differential reflectivity versus probe
delay at a wavelength close to the cavity resonance measured at a
different position than the cw reflectivity (Fig. \ref{fig:Refl}).
The width of the trough around $\Delta t = 0$ fs is indicative of
instantaneous probe absorption. (b) Differential reflectivity versus
probe delay and wavelength. At $\Delta t>100$ fs, $\Delta R/R$
increases (decreases) at the blue (red) edge of the cavity,
indicating a blue shift of the stopband and resonance. The black
squares are the extracted cavity resonance, connected by a guide to
the eye (red curve). The pump and probe energy are $E_{\rm{Pump}} =
1.8 \pm 0.18$ $\mu$J and $E_{\rm{Probe}} = 4 \pm 2$ nJ,
respectively.}
 \label{fig:3D}
 \end{center}
\end{figure}

We dynamically probe the excited cavity by pump-probe reflectivity.
Near pump and probe coincidence ($\Delta t = 0$ fs), the
differential reflectivity at the unswitched cavity resonance briefly
decreases during $\Delta \tau_0 = 218 \pm 5$ fs full width at half
minimum (FWHM), see Fig. \ref{fig:3D}a. This value agrees well with
$\sqrt{2} \tau_{\rm{P}} = 200 \pm 15$ fs for the cross-correlation
of the pump and probe pulses, which signals an instantaneous
non-linear process. Fig. \ref{fig:3D}a shows an increase of
reflectivity at longer probe delays. This is the result of the
excited free carriers \cite{Leonard:02, BeckerAPLetal:05} that
decrease the index and thereby blueshift the cavity resonance. After
about 50 ps, the changes in differential reflectivity have nearly
vanished due to the recombination of the free carriers.

To dynamically track the cavity resonance, we have measured the
time-resolved differential reflectivity for a large spectral
range, see Fig. \ref{fig:3D}b. The data clearly demonstrate the
free carrier induced blue shift: the differential reflectivity
increases at short wavelengths and decreases at long wavelengths.
From the data we have extracted the time-dependent cavity
resonance. \cite{notederive} Between 0 and 6 ps (=
$\tau_{\rm{ON}}$), the cavity resonance quickly shifts to shorter
wavelengths. The finite switching-on time is due to carrier
thermalization, and compares well to Ref.~\cite{Huang:98}. The
maximum shift is $\Delta \lambda = 5.6$ nm, corresponding to $3.3$
times the unswitched linewidth (\textit{cf.} Fig. \ref{fig:Refl}).
Subsequently, the wavelength of the resonance returns to the
unswitched case with a time constant $\tau_{\rm{OFF}}$ of about 50
ps. This direct observation of the ultrafast dynamic cavity
resonance shift and the subsequent relaxation is our main result.

\begin{figure}
\begin{center}
\includegraphics[angle=-90, width=8.0cm]{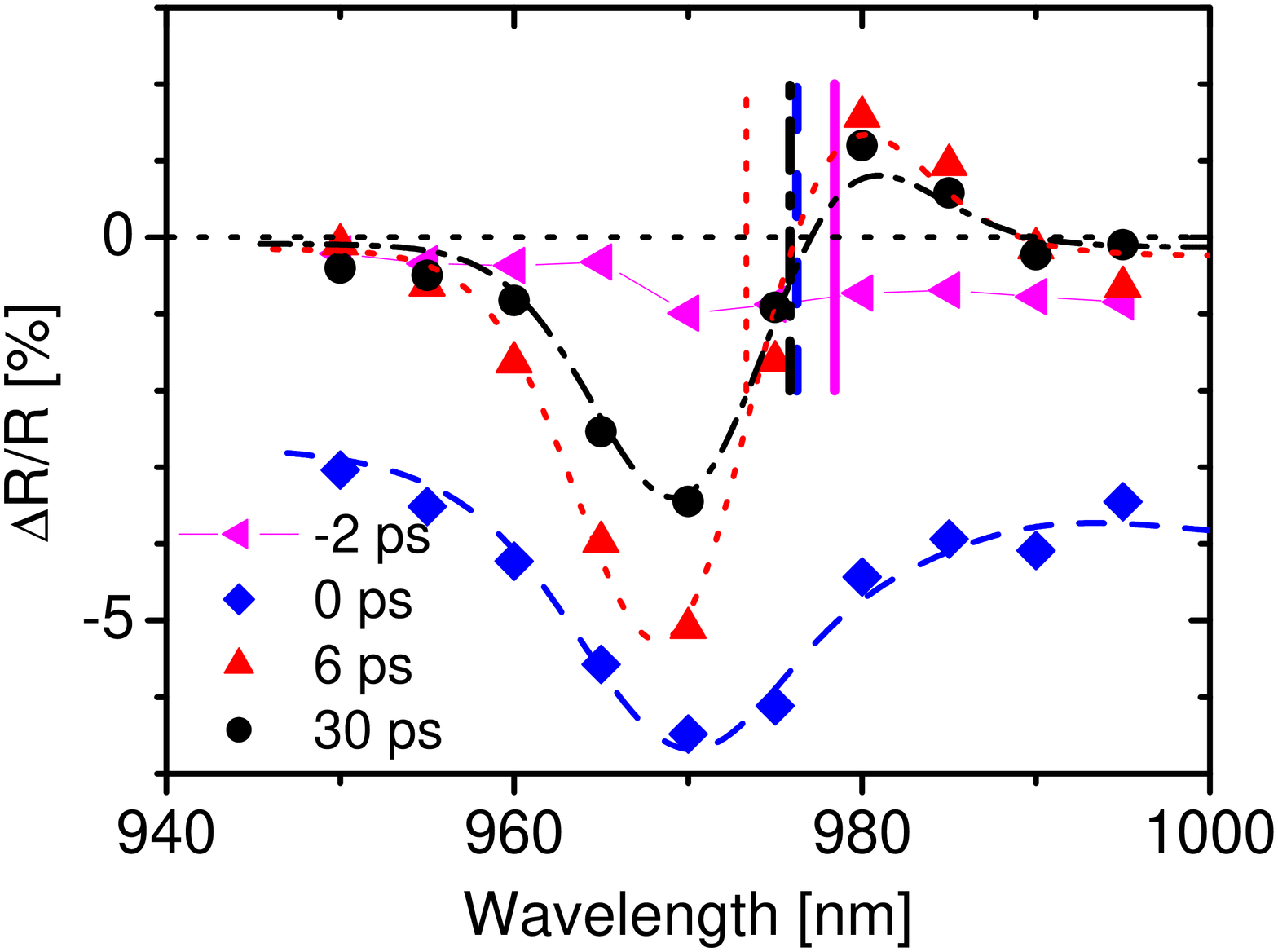}
\caption{(Color online) Differential reflectivity versus wavelength
for selected probe delays. The curves are transfer matrix
calculations, the vertical bars indicate the wavelengths of the
cavity resonance.}
 \label{fig:dRRvswavelength}
 \end{center}
\end{figure}

To investigate the dynamic behavior of the cavity in more detail, we
plot in Fig. \ref{fig:dRRvswavelength} the differential reflectivity
versus wavelength at selected delays. When the probe pulse arrives
before the pump pulse ($\Delta t<0$ ps), $\Delta R/R$ is slightly
negative since the polished rear side of the substrate reflects some
of probe pulse, which meets the pump on its way back where it gets
absorbed \cite{Euser:07a}. At pump and probe coincidence, the
differential reflectivity has decreased and reveals a broad minimum.
The decreased reflectivity is attributed to non-degenerate
two-photon absorption, since the sum of the pump and probe frequency
$E_{\rm{Total}} = 1.99$ eV is much above the optical bandgap of GaAs
(1.44 eV). At $\Delta t> 0$ ps, the differential reflectivity
acquires a dispersive shape, typical for the shift of a resonance.
Until $\Delta t = 6$ ps, the amplitude of the dispersive
differential reflectivity increases in magnitude, due to the
cavity's resonance shift, indicated by the bars in Fig.
\ref{fig:dRRvswavelength}. By interpreting the measured differential
reflectivity at $6$ ps with a TM calculation that includes a Drude
model to account for the excited carriers, we obtain a carrier
density of $N = 1.27 \pm 0.2 \times 10^{19}$ cm$^{-3}$.
\cite{notebeta}

\begin{figure}
\begin{center}
\includegraphics[angle=-90, width=8.0cm]{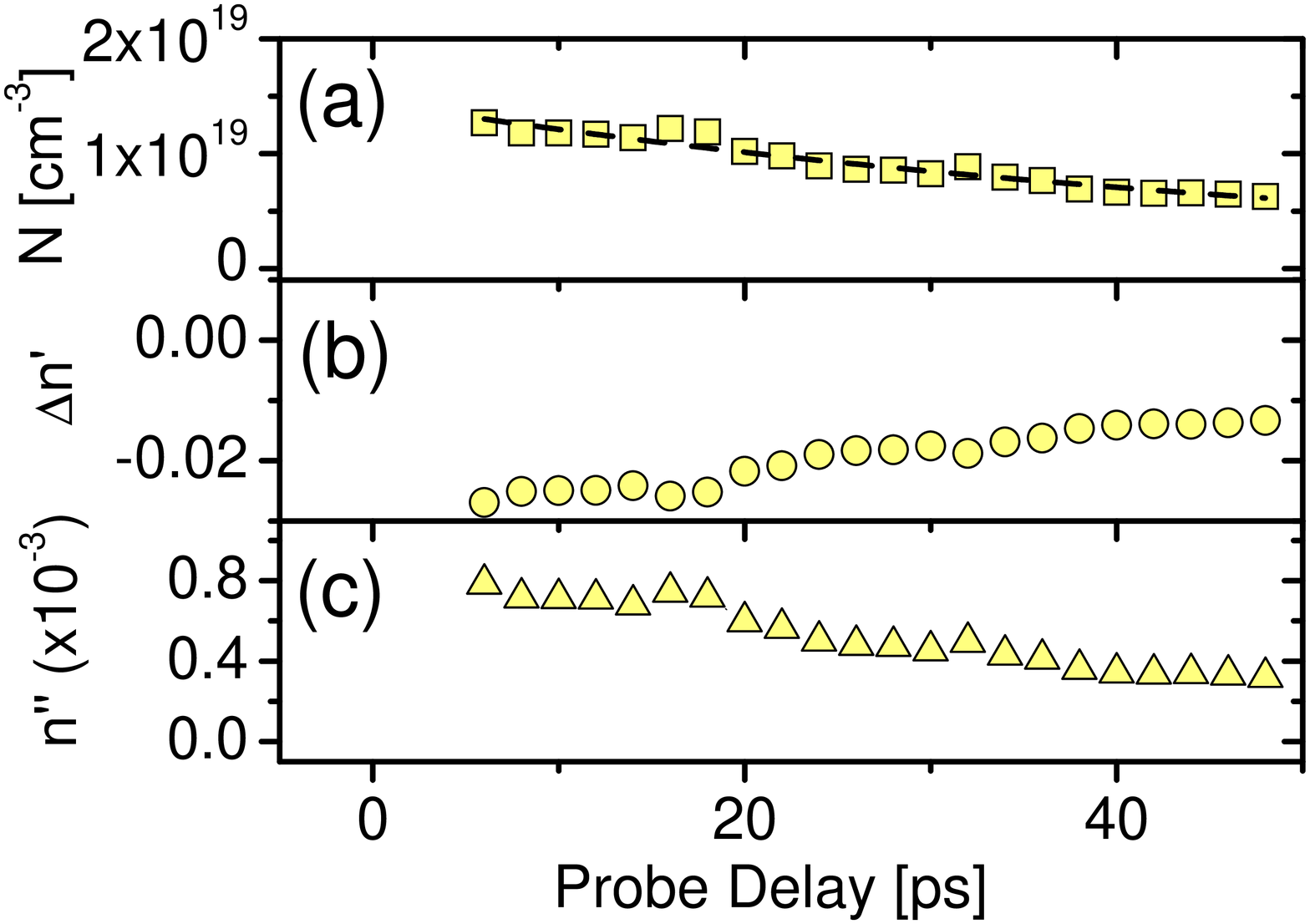}
\caption{(a) Carrier density versus probe delay as obtained from
the differential reflectivity using the extended Drude model at
delays of 6 ps and longer. We have fitted a single exponential
(dashed curve) to the carrier density with $\tau_{\rm{OFF}} = 55$
ps. (b) Change in real part n' and (c) imaginary part n'' of the
refractive index n calculated with the extended Drude model and
the carrier density. } \label{fig:DensityAndIOR}
\end{center}
\end{figure}

From the time- and wavelength resolved data, we obtain the dynamic
behavior of the carrier density $N$ shown in Fig.
\ref{fig:DensityAndIOR}(a). The Drude model was extended to include
electron-electron (e-e), electron-hole (e-h), and electron-phonon
scattering \cite{Hugel:99}. The only free parameter is a correction
to the Drude damping time accounting for e-e and e-h scattering, and
was found to be $10^5$ m/s. Absorption due to interband effects
appears to be only $1 \%$ of the free carrier absorption for our
experimental conditions. Using the Drude model before 6 ps is
unphysical as the electrons have not yet thermalized. After the
maximum density of $N=1.27 \times 10^{19}$ cm$^{-3}$ at $\Delta t =
6$ ps, the carrier density decreases with an exponential time
constant of 55 ps due to recombination. Thus, the total on-off cycle
can be accomplished in about $100$ ps (10 GHz), one order of
magnitude faster than previously reported \cite{Almeida:04}. The
maximum switching rates of microcavities may further be tenfold
increased \cite{Segschneider:97} by growing samples with a larger
number of recombination centers at the GaAs/AlAs interfaces.

From the free carrier density $N$ and the extended Drude model, we
have also calculated the time-dependent real (n') and imaginary
(n'') parts of the refractive index of the GaAs layers (Fig.
\ref{fig:DensityAndIOR}(b) and (c)). The real part mostly determines
the shift of the resonance wavelength, whereas the imaginary part
allows to assess possible changes of the quality factor $Q$ of the
cavity. The real part decreases by $\Delta n'_{\rm{GaAs}} = -0.027
\pm 0.004$, or $0.8\%$, corresponding to a 3.3 linewidth shift. The
imaginary part increases to $n''_{\rm{GaAs}} = 0.8 \times 10^{-3}$
due to the free carriers, before returning to the unswitched value.
From the maximum value of $n''$ at 6 ps, we estimate from a TM
calculation that Q has decreased to 220. Useful switching requires
the Bragg length $L_{\rm{B}}$ to be shorter than the absorption
length of either the pump ($\ell_{\rm{hom}})$ or probe
($\ell_{\rm{abs}}$) \cite{Euser:05, Euser:07a}; a natural figure of
merit (FOM) is then $(\ell_{\rm{hom}}^{-1} +
\ell_{\rm{abs}}^{-1})^{-1}/L_{\rm{B}}$. In our case, FOM$=22$,
comparable to Ref. \cite{Almeida:04} but much larger than Ref.
\cite{Fushman:07}, mostly due to their use of linear absorption.
Near $\Delta t=0$ fs, the imaginary index is briefly as large as
$n'' = 1.6 \pm 0.3 \times 10^{-2}$, corresponding to a decrease of Q
to 20. Here, $n''$ was obtained by fitting a TM calculation with a
complex $n$ to the measured differential reflectivity (Fig.
\ref{fig:dRRvswavelength}), and corresponds to a non-degenerate
two-photon absorption coefficient for GaAs of $\beta_{12} = 17 \pm
3$ cmGW$^{-1}$, in agreement with $\beta_{12} = 10$ cmGW$^{-1}$
derived from Ref. \cite{Hutchings:92}. While this period of
relatively high absorption lasts rather briefly, it is recommended
to keep the sum of the probe and pump frequencies below the optical
bandgap of the constituent materials or to reduce the probe and pump
fluences (see below).

In summary, we present unprecedented wavelength-resolved dynamic
behavior of a cavity resonance, with femtosecond resolution. Our
experiments were optimized for spatially homogeneous switching by
two-photon excitation to facilitate a physical interpretation of the
free carrier effects with an extended Drude model. Considerably
lower switching powers useful for future applications can be
realized by improving four features: 1) Using a much smaller pump
focus, a reduction in pump energy by a factor $\sim700$ can easily
be achieved. 2) Using one-photon absorption near the bandgap of GaAs
leads to a 100-fold reduced pump energy. Besides, it has been
predicted that a lower spatial homogeneity may be
favorable.\cite{Notomi:06} 3) Relaxing the shift of the cavity to
only one linewidth reduces the pump power by another factor of 3. 4)
Since the required pump energy scales inversely with Q, feasible
cavities with $Q \sim$50,000 \cite{Akahane:03} will reduce the pump
power by two orders of magnitude at the expense of the same
reduction in switching rate. Therefore, these simple consideration
already amount to a reduction of the pulse energies by a factor of
more than $2 \times 10^{7}$ to fJ, within reach of on-chip light
sources such as diode lasers.

We thank Henry van Driel and Allard Mosk for discussions. This
work is part of the research program of the "Stichting voor
Fundamenteel Onderzoek der Materie (FOM)", which was supported by
the "Nederlandse Organisatie voor Wetenschappelijk Onderzoek"
(NWO).

\begin{thebibliography}{}
\bibitem{notederive}
We have derived the wavelength of the cavity resonance by using the
identity
$$\frac{d(\Delta R/R)}{d \lambda} =
\frac{1}{R_0(\lambda)^{2}}\left(R_0(\lambda)\frac{d R(\Delta t,
\lambda)}{d \lambda} - R(\Delta t, \lambda)
\frac{dR_0(\lambda)}{d\lambda} \right) $$ where $R_0(\lambda)$ is
the unswitched reflectivity, and $R(\Delta t, \lambda)$ is the
switched reflectivity at delay $\Delta t$. Setting $\frac{d R(\Delta
t, \lambda)}{d \lambda} = 0$, which is the necessary extremal
condition for the cavity resonance, we obtain
$$  0 = \frac{d(\Delta R(\Delta t, \lambda)/R_0(\lambda))}{d
\lambda} + \frac{R(\Delta t, \lambda)}{R_0(\lambda)^2}
\frac{dR_0(\lambda)}{d\lambda}. $$ Thus, the dynamic resonance
wavelength is completely determined by the experimental data.
\bibitem{noteQD}
The maximum unbroadened refractive index change of the dots
amounts to only $10^{-8}$, while the absorption at resonance is
less than $0.02$ cm$^{-1}$.
\bibitem{notebeta}
From this density we infer a degenerate absorption coefficient of
$\beta = 0.53 \pm 0.11$ cmGW$^{-1}$.
\end{thebibliography}

\end{document}